# Current developments of nanoscale insight into corrosion protection by passive oxide films

Vincent Maurice, Philippe Marcus

*PSL Research University, CNRS - Chimie ParisTech,*
*Institut de Recherche de Chimie Paris / Physical Chemistry of*
*Surfaces Group*
*11 rue Pierre et Marie Curie, 75005 Paris, France*

Correspondence:

V. Maurice
E-mail: vincent.maurice@chimieparistech.psl.eu
Tel: +33(0)144278025
Fax: +33(0)146340753

P. Marcus
E-mail : philippe.marcus@chimieparistech.psl.eu
Tel: +33(0)144276738
Fax: +33(0)146340753

**Abstract**

Oxide passive films are a key for the durability of metals and alloys components as well as a central issue in corrosion science and engineering. Herein, we discuss current developments of the nanometer and sub-nanometer scale knowledge of the barrier properties and adsorption properties of passive oxide films brought by recent model experimental and theoretical investigations. The discussed aspects include (i) the chromium enrichment and its homogeneity at the nanoscale in passive films formed on Cr-bearing alloys such as stainless steel, (ii) the corrosion properties of grain boundaries in early intergranular corrosion before penetration and propagation in the grain boundary network, and (iii) the interaction of organic inhibitor molecules with incompletely passivated metallic surfaces. In all three cases, key issues are highlighted and future developments that we consider as most relevant are identified.

1. **Introduction**

Passive films provide the best of all means for corrosion protection of metallic materials. Thus passivity is a key for the durability of metals and alloys components, and a central issue in corrosion science and engineering. Passive films are oxides of the metal substrate, self-forming at the surface in aqueous environment and hydroxylated mostly in their outer part. The surface oxides are either thermodynamically stable or dissolve so slowly that they form an effective and well adherent barrier, most commonly only a few nanometers thick, between a corrosive environment and the substrate. Passive films are self-healing and may reform when locally degraded or removed by breakdown [1-8]. Self-healing is poisoned in the presence of aggressive anions (e.g. chlorides) which promotes the initiation of localized corrosion and thereby limits the durability of the self-protection provided to metallic materials [9-13].

Our current nanometer and sub-nanometer scale knowledge of the barrier properties of passive oxide films has been brought by combining a surface science approach employing single-crystal substrates with the application of surface analytical methods such as Scanning Tunneling Microscopy (STM), Atomic Force Microscopy (AFM) and Grazing Incidence X-ray Diffraction (GIXD) for characterizing the nanoscale morphology and the atomic structure of passive films produced under electrochemical control. Even when grown on single-crystal substrates, passive films are polycrystalline at the nanoscale, as a result of a high density of sites for oxide nucleation in their initial stages of formation, and textured with grains of nanometric lateral dimensions adopting a preferential crystallographic orientation. *Ex situ* as well as *in situ* measurements have demonstrated a crystalline atomic structure of the grains for passive films on metals such as copper [14-20], nickel [21-31], iron [32-39] and cobalt [40-42], and depending on the passivating conditions on chromium [43,44] and Cr-containing alloys like stainless steel [45-52] and nickel-based alloys [53]. The intergranular regions in between the nanograins may not form ordered grain boundaries with well-defined crystallographic character, in particular under non stationary conditions during the growth process. They act as preferential sites for local failure by passivity breakdown and the nanoscale initiation of localized corrosion [10,54-57].

Progress in our understanding of the atomic scale mechanisms by which the interaction of passive films with aggressive chlorides may lead to passivity breakdown has been brought by combining STM experiments with atomistic modeling using quantum chemical calculations. Density Functional Theory (DFT) modeling [58-60] suggested, in agreement with *in situ* STM data of dissolution in the passive state in the absence or presence of chlorides in the electrolyte [54-56], that the passive film may breakdown by a thinning mechanism promoted by chloride adsorption at step edges with formation of chloride- or hydroxychloride-containing dissolving metal complexes. For surface step edges of the passive film saturated by Cl adsorption, a penetration-induced mechanism of passivity breakdown may become prevalent. Reactive Molecular Dynamics (MD) calculations suggested that chloride ions preferentially interact with passive film surfaces that are deficient in oxygen [61,62].

In this article, we discuss current developments of the nanometer and sub-nanometer scale knowledge of the barrier properties and adsorption properties of passive oxide films brought by model experimental and theoretical investigations. Selected aspects that we think of paramount importance and insufficiently addressed until now are covered. Firstly, we discuss the deviation from homogeneity at the nanoscale of the chromium enrichment in passive films formed on Cr-bearing alloys such as stainless steel, a key issue for the durability of the self-protection of stainless metallic alloys in general. Secondly, we address the passivation property of grain boundaries of a metal emerging at the surface, also a key issue since most technical materials are polycrystals that expose a grain boundary network to the environment and primarily degrade by intergranular corrosion initiated at the nanoscale. Thirdly, we comment on recent data on the interaction of passivated metallic surfaces with organic inhibitor molecules, a high relevant issue in corrosion protection since inhibitors can be a major mean of protection in environmental conditions in which self-protection by passivation is incomplete and partially effective. Organic coatings and their adhesive interaction with oxide surfaces are beyond the scope of this paper since in this case corrosion protection is provided by the deposition of fully covering coatings and not by the self-protection behaviour of the metallic material by a passive film.

2. **Chromium enrichment in passive films on stainless steels**

Stainless steels and stainless alloys (e.g. nickel base alloys) are highly important technological materials owing their widespread use to the formation of markedly Cr(III)-enriched passive oxide layers, 1 to 3 nm thick at room temperature depending on passivation conditions and providing the high resistance to corrosion [1,2-8]. The chromium enrichment of the surface oxide formed in the passive range results from a small dissolution rate of Cr(III) compared to Fe(II)/Fe(III) oxides, with a difference even more effective for Cr enrichment in acid aqueous environment. Electrochemical overpotential for passivation and ageing time have been reported to markedly impact the crystallization of the passive film atomic structure as measured by STM on ferritic Fe-22Cr [47] and austenitic Fe-18Cr-13Ni [49,51] single-crystal alloys, and on Fe-Cr [45,46] and 304 stainless steels [48] sputter-deposited films, and concomitantly their dehydroxylation as measured by X-ray Photoelectron Spectroscopy (XPS) [47,49]. This is also the case on pure Cr [43,44]. The 2D atomic lattice observed by STM, an example is shown in Figure 1, was reported hexagonal with a parameter in the range 0.28-0.32 nm independently of the different substrates [45-49,51] and of the overall high Cr enrichment measured by XPS in the passive films [47,49]. It was attributed to the close-packed plane of the O sub-lattice of the Cr(III)-enriched nanocrystals developing in the oxide film concomitantly to dehydroxylation promoted by aging time and increasing passivation potential.

[INSERT Figure 1]

The similitude of the observed atomic lattice on the different pure Cr and stainless steel substrates is interesting. It suggests that the same O sub-lattice may host varying Cr(III) contents with no significant change measured by STM and, consequently, that the Cr(III) enrichment of the passive film on a given sample may not be homogeneous and vary locally as a result of inhomogeneities in the passivation process. This view is supported by the structural data compiled from Refs [63-,66] in Table 1 for chromium oxide ($\alpha$-$Cr_2O_3$), iron oxides (FeO, $Fe_3O_4$, $\gamma$- or $\alpha$-$Fe_2O_3$), nickel oxide (NiO) and mixed oxides ($FeCr_2O_4$, $NiCr_2O_4$). These oxides all have an O sub-lattice with a common hexagonal arrangement and anion nearest neighbor distance of 0.29-0.30 nm in the close packed planes, and it is only the distribution of the metal cations in this sub-lattice that specifies a given oxide structure.

The most recent nanoscale STM [51,52] and AFM [67] studies performed on stainless steel support this idea that the Cr enrichment in the passive film is not locally uniform. The STM images presented in Figure 2 displays the topography of a FeCrNiMo(100) single-crystal alloy surface passivated in sulfuric acid aqueous solution [52]. The terrace and step topography of the stainless steel substrate is clearly observed with a covering oxide film presenting a granular morphology at the nanoscale (Figure 2(a)). The lateral size of the grains was found to increase from 5.3 ± 0.9 nm for the native air-formed oxide film to 11.5 ± 2.6 nm after passivation by a potential step at 0.5 V/SHE in the middle of the passive range. In the absence of marked variation of the oxide film thickness, it was concluded that the initial surface oxide grains undergo a coalescence phenomenon accompanying the Cr enrichment induced by electrochemical passivation. Indeed, the overall oxide composition also changed during passivation with transient dissolution leading to higher Cr(III) enrichment as measured by XPS [52]: the Cr(III) and Fe(II)-Fe(III) cation fractions vary from 57 to 67% and from 41 to 29%, respectively.

[INSERT Figure 2]

The substrate terraces observed in Figure 2(a) also display local depressions with a penetration depth that increase from 1.02 ± 0.20 nm before electrochemical passivation to 2.27 ± 0.25 nm after polarization at 0.5 V/SHE, a value that exceeds the thickness of 1.9 nm measured for this passive oxide film [52]. These local topography variations are caused by substrate dissolution that competes transiently with oxide transformation during passivation. Hence dissolution as observed locally by STM can be concluded to be a marker of the surface regions of the passivated surface where the surface oxide was initially the least Cr-enriched.

The surface topography obtained after passivation also reveals that transient dissolution is not homogeneous at the nanoscale since it preferentially occurs on the terraces rather than at the step edges of the substrate. This is confirmed in Figure 2(b), showing a typical *in situ* EC-STM image recorded in the pre-passive range at -0.14 V/SHE, i.e. under polarization conditions promoting transient dissolution by decreasing the overpotential for formation of the passive oxide. In these conditions, depressions developed on the terraces and reached a penetration depth of 3.59 ± 0.66 nm after 72 min of polarization, deeper than those produced in Figure 2(a) after 120 min of polarization at higher overpotential (0.5 V/SHE). The preferential location of dissolution on the terraces may appear in contradiction with the nanoscale mechanism of active dissolution of metals taking place preferentially at step edges in the absence of a passivating oxide [68-70]. The difference arises from the competing passivation by Cr enrichment of the oxide film, most likely favored at step edges and thus promoting the protection of these substrate sites. Dissolution is then re-located on the substrate sites where the protection provided by the oxide film is less effective because of a lower local Cr-enrichment on the oxide film. Preferential passivation at step edges and roughening of substrate terraces by transient dissolution were also observed *in situ* for FeCrNi(100) passivated in similar conditions [51]. The fact that the terraces themselves do not dissolve homogenously but rather locally as evidenced in Figure 2(b) also shows that there are local sites on the terraces where the protection against dissolution fails to be effective due to local discontinuities in the Cr content of the oxide film.

The conductive AFM data presented in Figure 3 confirm that the passive film formed on stainless steel can be inhomogeneous at the nanoscale. They were obtained on 316L samples passivated in simulated PWR (pressurized water reactor) environment (pressurized water at 325°C) [67]. In these conditions of exposure at higher temperature, the passive film grows thicker (~5 nm after a 2 min exposure) and develop larger nanograins (29.1 ± 2.4 nm) than when formed at room temperature (Figure 3(a,c)). The grown oxide film is also less Cr(III)-enriched because of formation at near neutral pH and consists of a Cr(III)-rich inner barrier layer mixed with Fe(II,III) species and a mostly Fe(II,III)-rich outer layer.

[INSERT Figure 3]

Conductive AFM was applied to map the electrical resistance of the oxide film and thereby the homogeneity of its barrier properties. It was found that local electrical resistance measured on the oxide grains spread over ~1 order of magnitude with larger variations reaching 2–3 orders of magnitude observed locally (Figure 3(c,d)). The outer layer of the passive film being mostly constituted of Fe(II,III) oxide of low resistivity, the measured variations of the local resistance were assigned to the inhomogeneity of the resistivity of the Cr(III)-rich inner layer of the oxide film and its related composition. The grains having the highest resistance (resistivity) would consist of nearly pure Cr(III) ($Cr_2O_3$) oxide while those having a lower resistance (resistivity) would consist of Fe(II)-containing mixed (e.g. $FeCr_2O_4$) oxides, thus supporting the idea that the Cr enrichment varies between the nanograins themselves. Nanoscale variations of the surface electrical properties were also observed by conductive AFM on stainless steel fibers and assigned to the inhomogeneity of the resistivity, and thus barrier properties, of the air-formed surface oxide [71]

These studies show that passive oxide films formed on stainless steel in different environmental conditions exhibit a nanogranular morphology including intergranular sites, and that the local Cr(III) enrichment may vary depending on the coordination sites (steps vs. terraces) of the substrate and also between the oxide nanograins themselves. This provides new nanoscale insight to understand the local occurrence of passivity breakdown on Cr-bearing alloys. The less Cr(III)-enriched nanograins of the passive film, and even more the intergranular sites between such grains [10], are suggested to be the most susceptible to local breakdown owing to preferential iron dissolution in the passive state. Self-healing implies that the local composition in chromium in the modified alloy layer underneath the surface oxide in the breakdown sites remains high enough for local effective repassivation. Consequently, the local mechanisms by which metallic Cr is consumed in the topmost layers of the alloy by initial oxidation in ambient conditions and subsequently upon electrochemical passivation can be seen key issues for the nanoscale initiation of localized corrosion.

In our opinion, future developments on these aspects should include the detailed characterization by *in situ* spectroscopic and microscopic methods of the mechanisms of early oxidation of model stainless steel surfaces initially oxide-free in conditions simulating the formation of air-formed native oxides. The alterations of oxide films of well-controlled Cr enrichment prepared by oxidation under UHV (ultra high vacuum) environment by electrochemical polarization in varying overpotential conditions should also be studied. This requires to combine UHV facilities for surface preparation and characterization with *in situ* electrochemical testing of the corrosion properties and characterization of the corrosion-induced alterations and to ensure minimum or no alteration during transfer between UHV and liquid. Such studies would provide new comprehensive nanoscale knowledge that would serve as basis for designing surface treatments of stainless steels and stainless alloys for improved resistance to initiation of localized corrosion.

3. **Early stage intergranular corrosion**

Most technical metallic materials are polycrystals that expose to the environment a grain boundary network joining grains of different crystallographic orientations. Being microstructural defects that can have high energy depending on their structure, grain boundaries impact many properties of polycrystalline materials including their corrosion resistance [72,73]. Intergranular corrosion is a largely encountered form of localized corrosion for metals and alloys in the active state, i.e. when no or poorly protective passive films are formed, depending on the environmental conditions. Intergranular corrosion also occurs in the presence of a well passivating oxide film, for example in intergranular stress corrosion cracking [74]. With the development of nanocrystalline metallic materials [75], typically with a grain size lower than 100 nm and that have distinct properties from their coarse grains counterparts owing to the extremely fine grain size and the large volume fraction of interfaces, grain boundaries and triple junctions become surface sites of the uppermost importance for the corrosion resistance. Besides, the understanding of the early intergranular corrosion, i.e. in its initiation stage before it penetrates the sub-surface and propagates in the grain boundary network, becomes a requirement to control the degradation of materials of nanoscale dimensions (e.g. metallic interconnects for microelectronics).

Intergranular corrosion as observed by sub-surface penetrating attack of the grain boundary network has been shown to be intimately related to the crystallographic character and energy of the grain boundaries [72,74,76-90]. Low angle grain boundaries, having a misorientation angle < 15° and described by a network of misorientation dislocations (i.e steps) in the grain boundary plane, are considered as resistant to intergranular corrosion because of their lower energy compared to large angle grain boundaries. Among the large angle grain boundaries, coincidence site lattice (CSL) boundaries, labeled Σn with 1/n defining the density of lattice sites in common in the two grains, have been reported to better resist degradation than random boundaries, especially the low Σ ones with a high density of common lattice sites [72,77,79,82,83,84,86]. Among the low Σ CSLs, only Σ3 boundaries, which are the twins most commonly encountered in *fcc* materials, would better resist to intergranular corrosion [72,80,83,84,86], in particular the Σ3 coherent twins with in theory no deviation of the GB plane from the exact CSL {111} orientation [74,84].

Recent studies performed by EC-STM on high purity microcrystalline copper have demonstrated the possibility to investigate *in situ* at the nanoscale the initial stages of intergranular corrosion before penetration in the sub-surface region and propagation in the GB network [91] as well as the passivation properties [92,93] of grain boundaries emerging at the surface. A prerequisite to investigate the effect of the GB crystallographic character is the production of a high purity material, allowing us to exclude the segregation of impurities at grain boundaries that could impact the intergranular corrosion behavior. This was achieved in these studies using high purity cast electrolytic tough pitch (ETP-) Cu with trace contamination confirmed to be below the detection limit of Time-Of-Flight Secondary Ion Mass Spectrometry and cryogenic rolled and post-annealed at relatively low temperature (200°C) so as to limit grain boundary segregation and to keep a grain size compatible with the limited field of view of STM ($10 \times 10$ µm$^2$).

[INSERT Figure 4]

The EC-STM data presented in Figure 4 were obtained *in situ* on the oxide-free microcrystalline copper surface in chloridric acid aqueous solution [91]. They show that performing cycles of anodic dissolution in the absence of any surface oxide reveals a grain boundary type-dependent corrosion behavior, and thus how insightful the approach is to better understand early intergranular corrosion at the nanoscale. Two straight grain boundaries (marked by arrows) can be seen together with several curved grain boundaries, and their evolution after anodic dissolution is highlighted by comparative cross section analysis. A marked local attack is observed for the curved grain boundary (Figure 4d). The magnitude of intergranular corrosion is determined by both the depth penetration at the grain boundary and the dissolution at the surface of the grains that form the grain boundary. In this case, the depth penetration of ~2 nm, equivalent to ~10 Cu monolayers, was much higher than that (~0.1 nm) measured electrochemically over the whole electrode area from anodic charge transfer, showing that dissolution proceeds essentially at grain boundaries in the tested conditions. At the two straight grain boundaries, there is no indication of intergranular corrosion since both profiles are similar (Figure 4c). Electron back scatter diffraction (EBSD) analysis of the samples allowed to assign the straight grain boundaries to the category of coherent twins whereas the curved grain boundaries were either of random high angle type or non-coherent CSL (including Σ3s) boundaries.

These results obtained on microcrystalline copper in the active state of dissolution show that the initiation of intergranular corrosion measured at the nanoscale at grain boundaries emerging at the surface is dependent on the grain boundary character like observed in the sub-surface penetrating and propagation stages [72,74,80,83,84,86]. This original work was however limited by the absence of detailed characterization of the GB crystallographic character of the local sites where corrosion could be measured at the nanoscale. Current methodological developments to overcome this limitation include micro-marking of the surface by indentation with the STM tip after EC-STM analysis and subsequent repositioning of the EBSD analysis in the micro-marked local area [94]. The results obtained with this combined ECSTM/EBSD local approach show that, among the boundaries that could be locally analyzed, the surface-emerging random high angle boundaries are prone to nanoscale initiation of localized corrosion as well as Σ9 boundaries with a deviation of 0.6° from the exact CSL orientation. For Σ3 coincidence boundaries, the behavior of the emergence is found dependent on the deviation of the GB plane from the exact CSL orientation. Significant initiation of localized dissolution was observed for a deviation of 1.7° and above, whereas resistance to GB attack was observed for a deviation of 1° or less. It is thus shown that, in the early stages where only a few equivalent monolayers of material may corrode at the grain boundary emergences, increasing the density of steps (i.e. misorientation dislocations) in the coincidence plane of the boundary induces a transition from resistant to susceptible corrosion behavior. This transition is dependent on the CSL type.

Concerning the passive state, the passivation properties of grain boundaries emerging at the surface could be studied also on microcrystalline copper to get new insight onto grain boundary effects on the competition between irreversible consumption of the metal by dissolution and reversible consumption to form the passive film [92,93]. The adopted methodology included local examination by EC-STM of the surface in three distinct states: (i) the metallic state as obtained after cathodic reduction of the air-formed native oxide film, (ii) the Cu(I) or Cu(I)/Cu(II) passive state obtained by potentiostatic polarization in sodium hydroxide alkaline aqueous solutions and (iii) the metallic state obtained after cathodic reduction of the passive film. The results for the Cu(I) passive state are presented in Figure 5 [93].

[INSERT Figure 5]

The EC-STM image shows several curved grain boundaries and straight grain boundaries. The curve boundaries were assigned to either random or CSL (Σ3 incoherent twins or other Σn) boundaries and the straight grain boundaries to coherent twins. Different sites, marked 1 to 13, were selected for local analysis by line profile analysis of the depth of the intergranular surface region and its variation in the metallic, passivated and reduced surface states. The results compiled in Figure 5b show that the intergranular local topographical behavior is characterized by a decrease of the depth of the intergranular region after passivation for all sites, and a reversible variation after reduction of the passive film. This behavior suggests that locally (i.e. in the grain boundary region) a thicker Cu(I) passive film is formed than on the adjacent grains and that it is reversibly decomposed after electrochemical reduction, leaving the depth of the initial intergranular region (quasi) unchanged. The measured intergranular depth decrease was larger after passivation at the curved boundaries, suggesting the formation of a thicker Cu(I) passive film in the boundaries of random type.

A model of variation of the surface and interface levels at grains and grain boundaries in the studied metallic, passivated and reduced surface states was developed to exploit such local STM data together with the macroscopic electrochemical data obtained by cyclic voltammetry (CV) in terms of equivalent thickness of copper reversibly and irreversibly consumed at grains and grain boundaries [92,93]. The results are compiled in Figure 6a for the local grain boundary sites identified in Figure 5a. They show that the thickness of the Cu(I) passive film was systematically found larger at the selected grain boundary sites than macroscopically measured in average on grains by CV. These data also confirmed that the Cu(I) passive film is thicker at random boundaries than at coherent twins as inferred from the line profiles (Figure 5b). Consistently, the thickness of metallic copper reversibly consumed to form the passive film ($\delta_{Cu(R)}$) is also larger at grain boundaries than on grains. Concerning the equivalent thickness of metallic copper irreversibly consumed ($\delta_{Cu(IR)}$), i.e. dissolved, during passivation, the average value of 0.14 nm corresponding to a fraction of a monolayer of metallic copper was obtained for grains, from the macroscopic CV data.. For grain boundaries, the calculated values do not exceed that on grains. The slightly negative values obtained for some of the measured sites suggests that not all metallic copper reversibly consumed and entering the passive film was taken into account by the model. It could be concluded from these data that no copper is preferentially dissolved in the intergranular regions for Cu(I) passivation.

In the case of Cu(I)/Cu(II) passivation (Figure 6b) [92,93], the passive film was found thicker at random boundaries than at coherent twin boundaries like for the Cu(I) passive film. However, it was not found systematically thicker at grain boundaries than on grains, in contrast with the Cu(I) passive film. Preferential transient dissolution was observed in the intergranular regions during passivation, as shown by larger values of the equivalent thickness of metallic copper irreversibly consumed at grain boundaries compared to grains. A grain boundary crystallographic character-dependent behavior was observed with random boundaries dissolving more than coherent twins during passivation, in agreement with the behavior observed for active dissolution, i.e. in the absence of formation of a passive film [91].

These studies reveal how insightful the application of *in situ* EC-STM can be to investigate at the nanoscale and locally the initial stages of intergranular corrosion before sub-surface penetration and propagation in the GB network. Further developments should include combined EC-STM/EBSD characterization of the local passivation behavior of emerging grains boundaries in order to refine the characterization of the relationship between local passivation properties and grain boundary structure. Future work should include the effect of corrosive species (e.g. chloride) on the local passivation properties of different types of grain boundaries as well as the effect of corrosion inhibitor molecules (e.g. BTAH for copper) on the intergranular corrosion initiation in the active state and in the passive state.

4. **Atomistic modeling of corrosion protection on passivated surfaces by organic molecules**

Molecular additives, inorganic or organic, are commonly introduced as inhibitors in the environment for decreasing the rate of the corrosion processes of materials in conditions in which passivation is dysfunctional. In order to be efficient, the inhibitor molecule must adsorb on the material surface, the adsorption must be strong, stronger than that of the corroding species, and ideally a barrier layer should form isolating the metal substrate from the environment and it should not be displaced by solvent molecules and/or aggressive species.

On the macroscopic scale, organic corrosion inhibitors are mostly studied experimentally with information gathered during sample immersion and/or after immersion to bring insight into the effectiveness of different chemicals. However, the relevant scale to understand the inhibiting effect and to seek details for a better design of new and environmental-friendly inhibition systems is the molecular scale that requires a realistic model approach. Experimental studies of corrosion inhibition are numerous [95-99] but the development of atomistic theoretical investigations is more recent [100,101]. Computational model studies bring detailed information on the atomic and sub-atomic scale on the adsorption of the organic inhibitor molecule from the structural, energetic and electronic point of views. They can also be used to screen much more systems and effects than it can be done experimentally, however with the disadvantage of lacking counterpart model studies performed experimentally at the same scale. Even for the benzotriazole/Cu system, the most studied both experimentally [102-109] and computationally [110-113] with model approaches, experiments and modeling have not succeeded in describing the system with the same degree of complexity. Atomistic DFT modeling has addressed the effects of the molecule coverage, its deprotonation, the coordination of the metal surface atoms, the presence of surface metal adatoms, the competitive adsorption of water, and the type of method used for computation while experimental model studies at the molecular scale have only discussed the effect of the molecule coverage on the structure and adsorption strength of the adsorbed phases [108,109], and the displacement of the adsorbed molecular layer by aggressive species [102,104].

The role played by the surface oxide film formed on passivated copper on the adsorption of benzotriazole (BTAH) has been recently studied by atomistic DFT modeling on (111)-oriented $Cu_2O$ surfaces [110-112]. Copper oxides are indeed formed on copper exposed to aqueous solution and do not dissolve at pH > 5, which results in surface passivation by (111)-oriented $Cu_2O$ oxide films on (111)-oriented copper in the Cu(I) oxidation range [16,17,18,20]. It was found that, like on oxide-free metallic copper, the benzotriazole molecules binds to the Cu sites of the $Cu_2O$ surface *via* the triazole group and its N atoms (Figure 7) [111]. The protonated BTAH molecule forms up to two bonds with Cu atoms (Figure 7(a,b,c,f,g)) and deprotonated BTA at least two (Figure 7(d,e,h,i)). There is relatively strong H-bonding for BTAH to one O of the oxide surface. Another common feature with metallic copper is that the deprotonated BTA binds markedly stronger to the oxide surface than BTAH and that bond strength increases when involving Cu sites that are coordinatively unsaturated. BTA binding energy is -2.8 eV/molecule when coordinatively unsaturated Cu sites of stoichiometric $Cu_2O(111)$ are involved (Figure 7(h,i)) against about -1.9 eV/molecule when only involving the coordinatively saturated sites of non-stoichiometric $Cu_2O(111)$ (i.e. without Cu unsaturated termination) (Figure 7(d,e)). For BTAH, these values are -1.5 and -0.4 eV/molecule, respectively. The energetic deficiency of the Cu unsaturated termination of stoichiometric $Cu_2O(111)$ was calculated to be overcompensated by the bonding of benzotriazole. Van der Waals interactions were implemented with PBE-D' functionals in DFT to better include dispersion forces in the optimization of protonated and deprotonated adsorbates and to better account for coverage effects. As a result, the adsorbed structures were stabilized by about 0.3-0.4 eV mol$^{-1}$ on average and the relative stability trends were unchanged (Figure 7).

[INSERT Figure 7]

Comparing the stability trends of the adsorbed molecules shows that the protonated BTAH molecule binds markedly stronger to Cu sites of reduced coordination on Cu$_2$O than on metallic copper, whereas this trend is reversed for the deprotonated BTA molecule. BTA bonding to the oxide requires coordinatively unsaturated sites in order to be comparably strong to the bonding to the coordinatively saturated sites on the metal. However, under-coordinated Cu sites on close-packed metallic surfaces or Cu sites of more open Cu(hkl) form stronger bonds to BTA than coordinatively unsaturated Cu sites on Cu$_2$O.

The effects of surface hydroxylation, relevant of the oxide/liquid water interface, were also modelled [112]. It was found that BTA binds considerably stronger to the hydroxylated Cu$_2$O surface (-3.8 eV/molecule) than to the anhydrous surface when involving Cu sites that are coordinatively unsaturated whereas the bond strength is similar (about -1.0 eV/molecule) for BTAH, showing that deprotonation is promoted by surface hydroxylation of the oxide. Benzotriazole forms more stable hydrogen bonds with the surface hydroxyls than with the O ions of the anhydrous surface which was assigned to the higher configurational flexibility of the surface OH groups. The deprotonation of the BTAH molecule leads to the formation of a water molecule on the hydroxylated surface in contrast to a hydroxyl group on the anhydrous surface.

Based on these results, we rationalize the efficiency of the BTAH inhibitor depending on pH of the aqueous solution and on the potential at the interface as follows. Protonated BTAH is the major form of benzotriazole at neutral and acidic pH since the molecule has a pKa constant of 8.2 [111]. In acidic conditions (pH < 5), benzotriazole is known as a less effective inhibitor than in alkaline conditions and copper does not grow a stable and protective surface oxide. The lower efficiency of benzotriazole can be explained by the DFT results that show that the protonated BTAH molecule has a lower adsorption energy than its deprotonated BTA variant and thus a higher probability for desorption. As a result, the barrier layer formed by the adsorbed molecules may become locally defective where the molecules were the less strongly bonded initially, and this local failure cannot be compensated by the formation of a passivating surface oxide thus leaving unprotected site for corrosive attack. In neutral or alkaline conditions (pH >5), passivating Cu oxide films grow in the anodic oxidation range. At potentials below the Nernst oxidation potential, the oxide-free metallic surface may be covered by the deprotonated BTA, more strongly adsorbed according to the DFT data, if pH is alkaline enough (> 8). At the Nernst oxidation potential, the growth of the passive oxide film can be expected to be poisoned or even blocked by the strongly adsorbed BTA layer. However, if this layer has barrier defects due for instance to weaker bonding of BTAH for 5 < pH < 8, Cu oxide may grow locally to form oxide patches covered by BTAH that can be strongly adsorbed according to DFT. In these conditions, the efficiency of the inhibitor results from its capacity to strongly adsorbed different variants according to the oxidized or metallic local state of the surface and to local pH conditions. However, the boundary between oxide patches and metallic areas exposes an oxide/metal interface where the inhibitor may be less or not effective if no variant of the molecule strongly adsorbs, which remains to be investigated with appropriate model approaches. At higher oversaturation potential for anodic oxidation, the strongly adsorbed BTA on the metallic surface is displaced and anodic oxides will form unless their growth is poisoned by aggressive species like Cl. The efficiency of the inhibitor is then dependent on its role on the competitive adsorption between Cl and $H_2O$/OH, the latter being the oxygen reactant for oxide growth in neutral/alkaline conditions. This also remains to be studied by realistic modeling of competitive adsorption processes.

Another approach to the DFT modeling of the effect of inhibitor organic molecules on the corrosion properties of metallic surface is to consider their role on the residual reactivity of passivated surfaces in open circuit potential (OCP) conditions. Indeed, in the absence of any applied potential, the anodic oxidation reactions at the origin of the corrosion processes are electrically balanced by cathodic reduction reactions. In neutral aqueous environment, the dominant cathodic reaction is the oxygen reduction reaction (ORR). Controlling this reaction is also a mean of corrosion protection, including for passivated surfaces. Atomistic DFT modeling has recently been applied to study this reaction in the presence of a model passive oxide film [114] and the inhibiting effect of adsorbed organic molecules has been discussed [115]. The work was performed on a (111)-oriented Al surface covered by an ultrathin oxide film ($\gamma$-$Al_2O_3$) of varying thickness and exposing an hydroxylated AlOOH surface developed as a realistic model of the passive film formed in neutral solution on Al and Al alloys. Figure 8a shows the ultrathin variant of the model where the oxide film is ~0.2 nm thick and consists of two layer of oxygen atoms. Periodic DFT including dispersion forces (PBE-D) was used for simulation.

In the absence of adsorbed organic molecule [114], it was found that electron transfer takes place both from the metal/oxide interface and the oxide surface to the adsorbed $O_2$ molecule. The hydroxyl groups at the oxide surface were found to stabilize the surface by eliminating the Al surface states and to allow the reduced $O_2^-$ species to transform into hydrogen peroxide in a non-activated process by proton capture. A series of two-electron mechanisms followed to further reduce $H_2O_2$ to two water molecules. These reactions were found to take place only when the inner alumina layer was ultrathin (0.2 nm in the studied model). As soon as a thicker $Al_2O_3$ inner layer of the surface oxide developed (for a calculated film thickness of about 1 nm), it was found that the electronic transfer from the metal/oxide interface to the oxygen molecule approaching the hydroxylated surface of the oxide was shut down and the oxide-covered surface became unreactive towards oxygen reduction. Thus, although a surface oxide layer can be expected to have a protective character against the cathodic reduction of $O_2$, it was shown in this work that $O_2$ reduction is not inhibited when the oxide film is ultrathin. The ultrathin oxide layer represents a possible model for incomplete passivation in conditions where the passive film dissolves as can be expected for amphoretic alumina in acidic or alkaline aqueous environment.

[INSERT Figure 8]

The effect of the adsorption of an organic inhibitor molecule and its ability to form a dense, continuous layer on the ultrathin oxide film surface and be protective against $O_2$ reduction was studied with gallic acid [115]. Among carboxylates, gallic acid (Figure 8b), reported to have an inhibiting effect on Al [116] and steel [117], was selected since it is the smallest molecule in the family of tannins which are green inhibitor candidates. A two-step strategy was used for the optimization of (i) a compact layer of free molecules adopting the registry of the model passivated surface and (ii) the bonding of this molecular layer to the hydroxylated surface of the oxide ultrathin film as detailed in [115]. It was found that the organic molecule is preferentially adsorbed in a monodentate way by ligand exchange, with an adsorption energy of 2.3 eV/molecule, and form a dense adsorbed layer with a density of 3.6 molecules/nm$^2$ (Figure 8c). The electronic analysis showed that covalent bonding was at the origin of the surface stabilization due to mixing of the O 2p orbitals of the COOH moiety of the molecule with the Al 2p orbitals forming the valence band of the oxide surface (Figure 8e). The metallic character of the ultrathin surface oxide observed in the absence of gallic acid was lost in the presence of the adsorbed dense layer with the oxide recovering a semi-conducting character as revealed by a marked decreased of the density of states of the oxide in the gap at the Fermi level (Figure 8d) [115]. Permittivity calculations based on charge analysis showed that a huge drop from a value of 109 in the absence of adsorbed gallic acid to 4 for the adsorbed system (a value of 9 was calculated for bulk alumina), confirming the much more insulating character of the surface covered by the dense adsorbed layer. These results suggest that adsorption of an inhibitor molecule would enable a passivating alumina film to recover its insulating properties at defective zones where it would be thinner, thus "healing" the defective site.

The dense adsorbed organic layer was confirmed to form an efficient barrier against electron transfer to dioxygen, as shown by the analysis of the O-O distance in the $O_2$ molecule and of electronic transfer. In the absence of adsorbed gallic acid, the O−O bond length spontaneously increased when the dioxygen molecule was at the vicinity of the surface, due to the transfer of one electron from the metal layers underneath the oxide film to the molecule. The charged $OO^-$ molecule then attracted a proton from the surface to form a OOH adduct. With the dense adsorbed layer of gallic acid, the O-O distance in the dioxygen molecule was found unchanged (0.127 nm) and the charge transfer was only 0.32 electron per $O_2$ molecule. No OOH adduct was formed. This altered behavior of the adsorbed system suggests good inhibition of the cathodic reaction by the adsorbed layer.

These studies show how powerful the application of DFT modeling can be to rationalize the mechanisms by which adsorbed organic molecules can inhibit the corrosion of incompletely passivated metal surfaces. The data simulating the adsorption of benzotriazole on copper show the capability of the molecule to efficiently comply with oxidized as well as metallic states of the surface and to provide strong bonding in the protonated BTAH and deprotonated BTA form, respectively, with stronger bonding to unsaturated (i.e. defect) Cu sites both on $Cu_2O$ oxide surfaces and on Cu metal. Still, one lacks similar studies on surfaces partially oxidized and exposing oxide/metal interfaces The data simulating the adsorption of gallic acid on passivated aluminium show that it is essential to take into account the presence of a surface oxide of finite thickness, and not only the surface of the bulk oxide, to properly describe inhibition on passivated surfaces. It comes out of these studies that adhesion of the organic molecule and stabilization of the electronic levels of the passivated surface can result at least partly from strong anchoring of the adsorbed organic molecule to the surface due to covalent bonding. In the presence of the dense layer of adsorbed molecules, the electronic states of the ultrathin oxide can become similar to those of the thicker oxide, inhibiting electronic transfer and thus cathodic corrosion processes. It is our opinion that future developments of DFT simulation for the detailed understanding of corrosion inhibition brought by organic molecules should consider surfaces exposing oxide/metal interfaces as well as oxide films of varying thickness in order to better account for the effects of incomplete passivation.

5. **Conclusion**

Three aspects that are, in our opinion, of paramount importance for the development of the nanoscale and sub-nanoscale understanding of the corrosion and corrosion protection mechanisms of metallic materials by oxide passive films have been discussed. The discussion was based on recent data obtained from model experimental studies, including local surface analysis by scanning probe microscopies, and atomistic modeling and simulation performed by DFT.

The discussed data on the passivation of stainless steel show that the Cr enrichment of the passive film may not be homogeneous at the nanoscale and vary between the oxide nanograins constituting the barrier layer of the passive film. This new nanoscale insight implies that the local mechanisms by which metallic Cr is consumed in the topmost layers of the alloy by initial oxidation and subsequently upon electrochemical passivation are key factors for passivity breakdown and self-healing, and for the initiation of localized corrosion. In our opinion, future work should include the detailed characterization of these mechanisms by *in situ* spectroscopic and microscopic methods on appropriately designed model systems. This approach will provide new comprehensive knowledge that can be used for the desihn of novel surface treatments for improved resistance to initiation of localized corrosion of stainless steel.

The discussed data on early intergranular corrosion of microcrystalline copper show that nanoscale insight into the relationship between grain boundary crystallographic character and local corrosion properties in the active or passive state can by seek by combining surface *in situ* analysis and microstructural analysis. The data show that $\Sigma 3$ coherent twins also resist the initiation of intergranular corrosion, the initiation being here the stage that precedes sub-surface penetration and propagation in the grain boundary network. Future work, aiming at improving the design of corrosion resistant grain boundary engineered materials, should include the effect of corrosive species such as chlorides on the local passivation properties of different types of grain boundaries as well as the effect of corrosion inhibitor molecules on the intergranular corrosion initiation in both the active state and the passive state.

The discussed data on the DFT modeling of corrosion inhibition by adsorbed organic molecules of the corrosion of incompletely passivated metal surfaces show the necessity for the molecule to efficiently comply with and strongly bond to surfaces in the oxidized as well as metallic states. They also show that it is essential to take into account the presence of a surface oxide of finite thickness, and not only the surface of the bulk oxide, to properly describe corrosion inhibition of passivated surfaces. In addition to oxide films of varying thickness or a thick oxide layer, future work by DFT modeling should also include partially oxidized metal interfaces in order to better account for the effects of incomplete passivation.

**Acknowledgments**

This project has received funding from the European Research Council (ERC) under the European Union's Horizon 2020 research and innovation programme (ERC Advanced Grant "CIMNAS" No 741123).

**Figure captions**

Figure 1. Atomic structure as observed *in situ* by EC-STM on Fe-18Cr-13Ni(100) under polarization in the passive range at 0.5 V/SHE in 0.5 M $H_2SO_4$(aq). The atomic lattice terminating the grains of the passive oxide film is hexagonal as shown by the inset. Adapted from [51].

Figure 2. Nanoscale morphology of the Fe-17Cr-14.5Ni-2.3Mo(100) stainless steel surface passivated in 0.05 M $H_2SO_4$(aq) as observed by STM after polarization at 0.5 V/SHE (a) and by EC-STM under polarization at -0.14 V/SHE. The passive oxide film has a nanogranular morphology resolved in (b) (some grains are pointed) and covering the substrate terraces and step edges (marked by dashed lines). Substrate terraces display depressions (some are circled in (a) and (b)) evidencing local protection failure caused by competing transient dissolution during passivation. Polarization in the pre-passive range in (b) causes the growth of the depressions by sustained transient dissolution. Substrate step edges (marked by dashed lines) are more corrosion resistant owing to preferential local Cr enrichment of the passive film. (a) Adapted from [52].

Figure 3. Nanoscale morphology (a) and electrical resistance map (c) as observed by current sensing AFM of the passive oxide film formed on 316L stainless steel exposed 2 min to simulated pressurized water reactor environment (water at 325°C). Height (b) and resistance (d) profiles along the green and white arrows in (a) and (c), respectively. Reproduced from [67].

Figure 4. Intergranular active dissolution as observed *in situ* by EC-STM on microcrystalline copper in 1 mM HCl(aq.). Topographic images before (a) and after (b) anodic dissolution and height profiles along line 1 (c) and line 2 (d). The arrows mark coherent twin grain boundaries and the rectangle marks a random grain boundary. Reproduced from [91].

Figure 5. Intergranular passivation as studied *in situ* by EC-STM on microcrystalline copper in 0.1 M NaOH(aq.). (a) Topographic STM image of the analyzed local area. The local sites selected for GB depth analysis across random and coherent twin boundaries are marked 1 to 4 and 5 to 13, respectively. (b) Bar graph of the GB depth measured in the metallic, passivated and reduced surface states at the marked selected sites in (a). Adapted from [93].

Figure 6. Bar graphs for the Cu(I) (a) and Cu(I/Cu(II) (b) passivation properties measured at grain surfacess (G) and at random (RGB) and coherent twin (CTGB) grain boundaries of microcrystalline copper in 0.1 M NaOH(aq.). The thickness of the passive film is displayed as the sum of the thickness of copper reversibly consumed by the passivation/reduction treatment ($\delta_{Cu(R)}$), the thickness of metallic copper irreversibly consumed (i.e. dissolved) during passivation ($\delta_{Cu(IR)}$), and the rise of surface level caused by passivation ($\Delta z_{Pass}$). Adapted from [93].

Figure 7. DFT optimized structures and binding energies of molecularly adsorbed benzotriazole BTAH (left) and deprotonated BTA (right) on (111)-oriented $Cu_2O$: (a-e) bonding to the non-stoichiometric termination without the Cu unsaturated sites (labelled CUS), (f-i) bonding to the coordinatively saturated sites (labelled CSA) and CUS sites of the stoichiometric termination. H-bonding forms between BTAH and surface oxygen (b,c,f,g). Binding energies (eV/molecule) as calculated with PBE and PBE-D' functionals are given as plain and bracketed values, respectively. Reproduced from [111].

Figure 8. DFT optimized structures and projected density of states (DOS) of molecularly adsorbed gallic acid on model passivated Al(111) surface. (a) Model of the ultrathin (0.2 nm) supported oxide film on Al(111). Topmost surface is hydroxylated. (b) Gallic acid molecule. (c) Dense adsorbed layer formed by the molecule on the supported oxide film. (d) Projected DOS of the oxide layer without (dotted line) and with (plain line) adsorbed layer. (e) Projected DOS of the adsorbed molecule COO function (gray line) and oxide layer (plain line). Zero energy level is set at the metal Fermi level. Adapted from [115].

**Tables**

Table 1. Structure type and lattice constants of various Cr-, Fe- and/or Ni-containing oxide compounds and of their oxygen sub-lattice, and orientation and nearest neighbor distance in the closed packed plane of the oxygen sub-lattice.

| Oxide | Ref. | Oxide compound | O sub-lattice | O close packed plane |
|---|---|---|---|---|
| α-$Cr_2O_3$ | 63 | Corundum<br>a = 0.496 nm<br>c = 1.360 nm | hcp<br>$a_O$ = 0.496 nm<br>$c_O$ = 1.360 nm | (0001)<br>0.286 nm |
| FeO | 64 | Rock salt<br>a = 0.430 nm | cfc<br>$a_O$ = 0.430 nm | {111}<br>0.304 nm |
| $Fe_3O_4$ | 64 | Inverse spinel<br>a = 0.839 nm | cfc<br>$a_O$ = 0.420 nm | {111}<br>0.297 nm |
| α-$Fe_2O_3$ | 64 | Corundum<br>a = 0.504 nm<br>c = 1.375 nm | hcp<br>$a_O$ = 0.504 nm<br>$c_O$ = 1.375 nm | (0001)<br>0.291 nm |
| γ-$Fe_2O_3$ | 64 | Defect spinel<br>a = 0.835 nm | cfc<br>$a_O$ = 0.417 nm | {111}<br>0.295 nm |
| NiO | 65 | Rock salt<br>a = 0.418 nm | cfc<br>$a_O$ = 0.418 nm | {111}<br>0.295 nm |
| $FeCr_2O_4$ | 66 | Spinel<br>a = 0.838 nm | cfc<br>$a_O$ = 0.419 nm | {111}<br>0.296 nm |
| $NiCr_2O_4$ | 66 | Spinel<br>a = 0.832 nm | cfc<br>$a_O$ = 0.417 nm | {111}<br>0.294 nm |

**Figure 1**

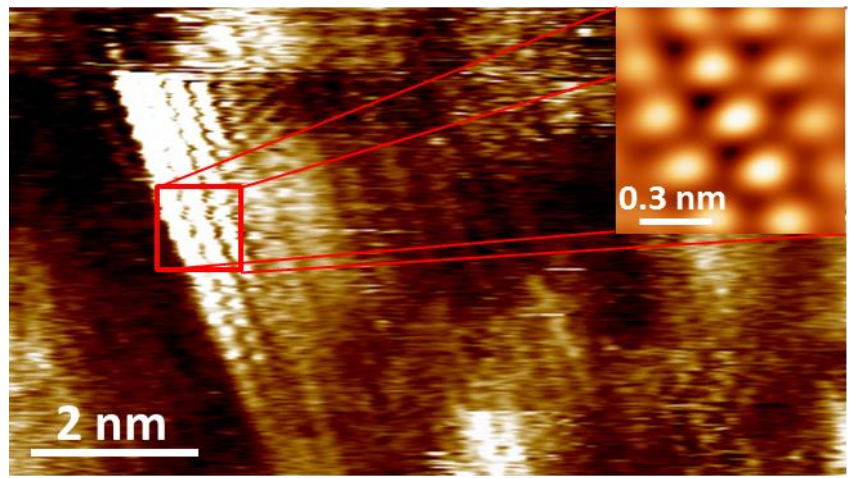

**Figure 2**

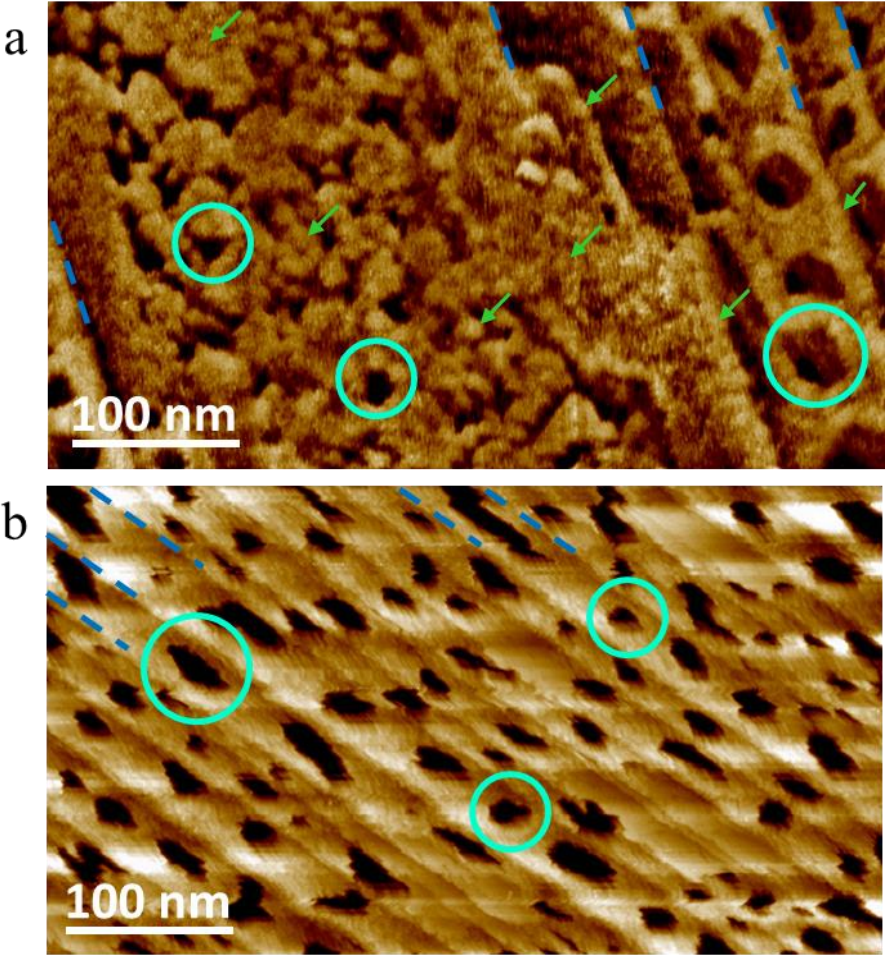

**Figure 3**

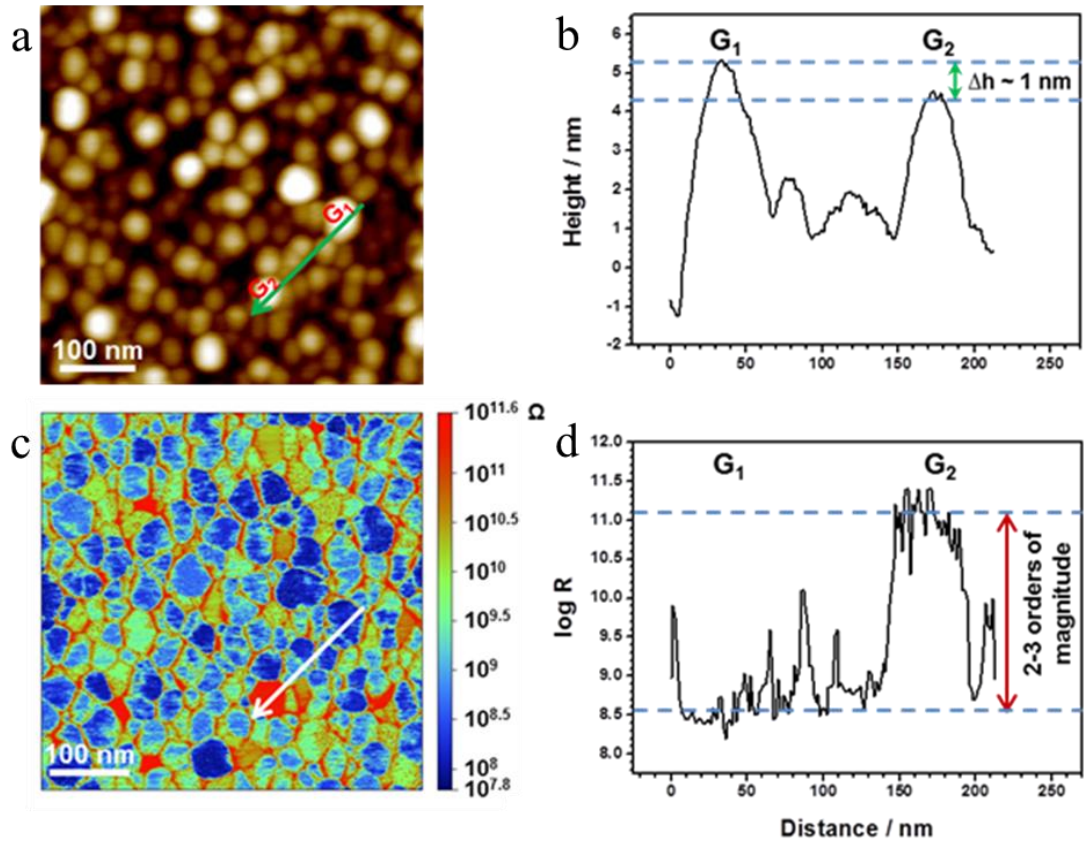

**Figure 3**

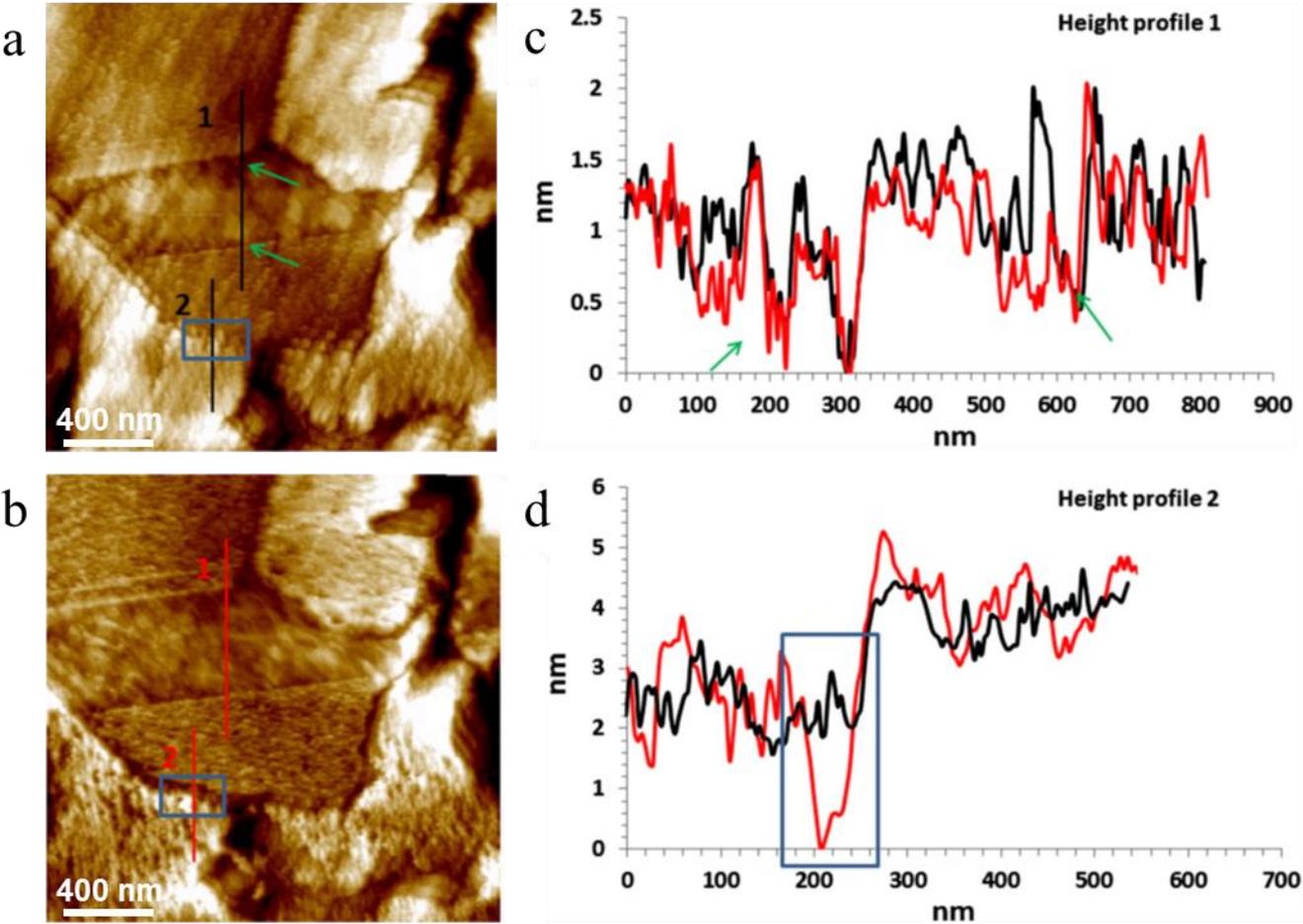

**Figure 5**

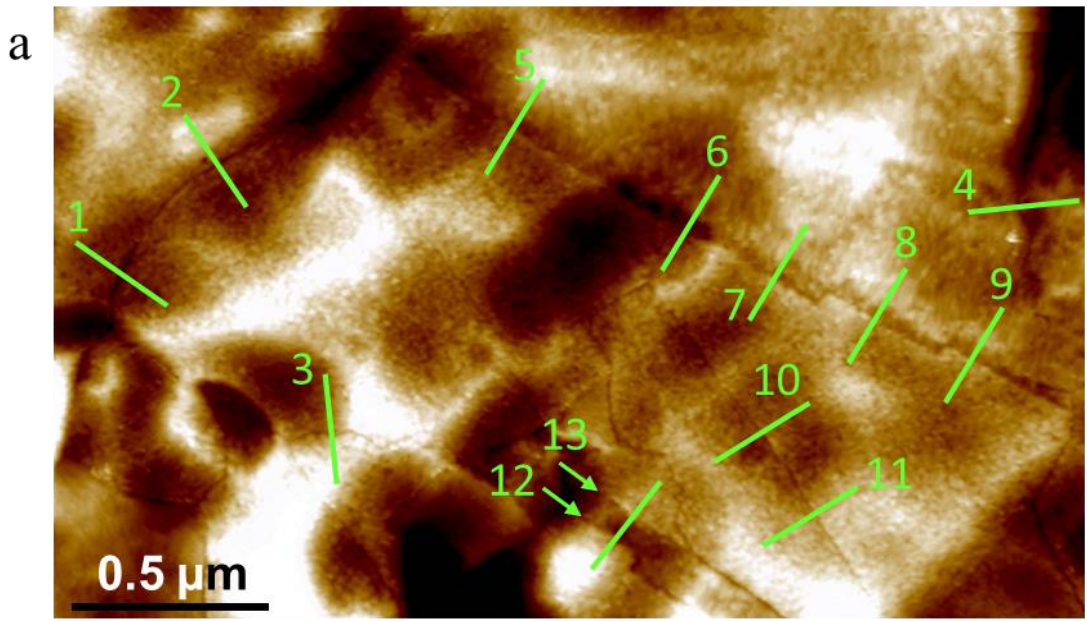

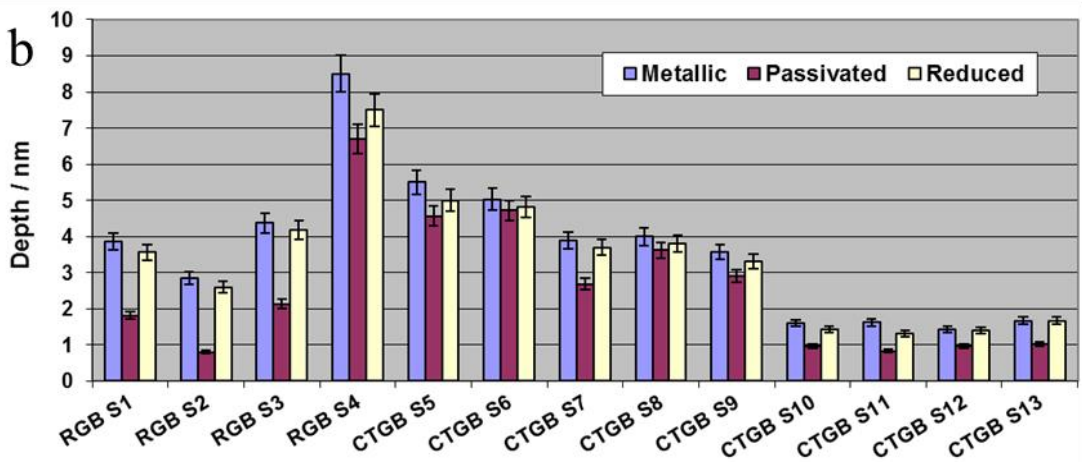

**Figure 6**

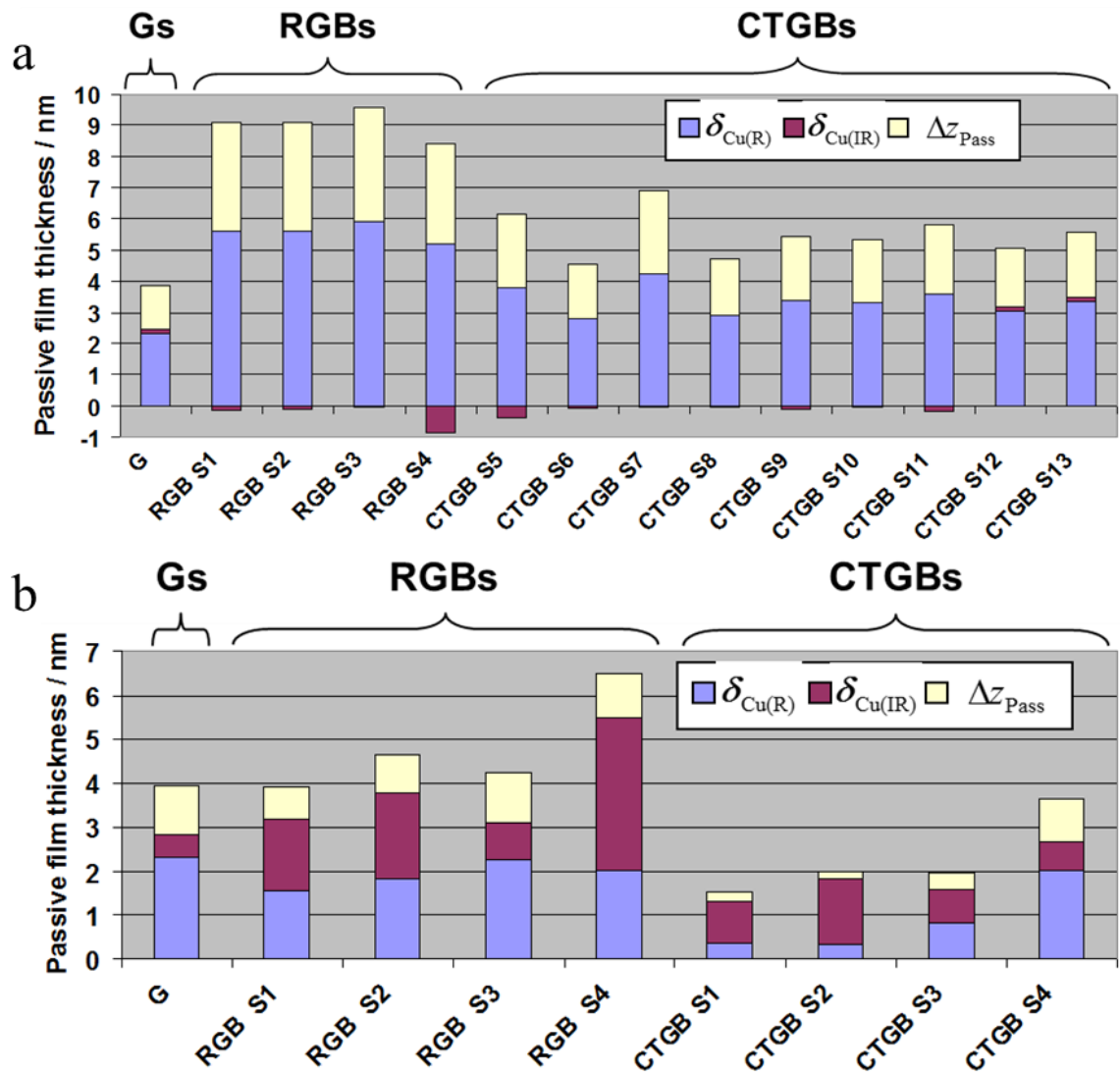

**Figure 7**

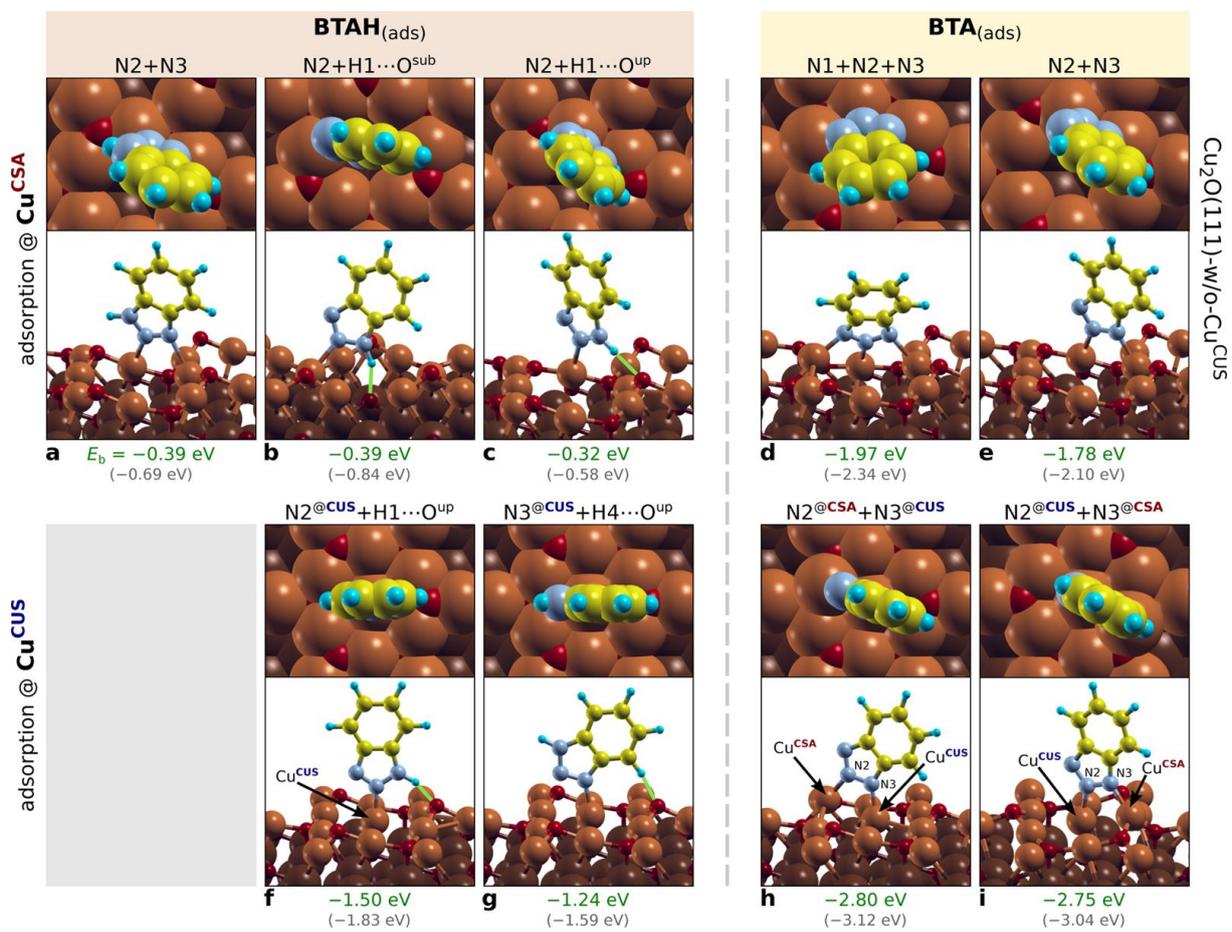

**Figure 8**

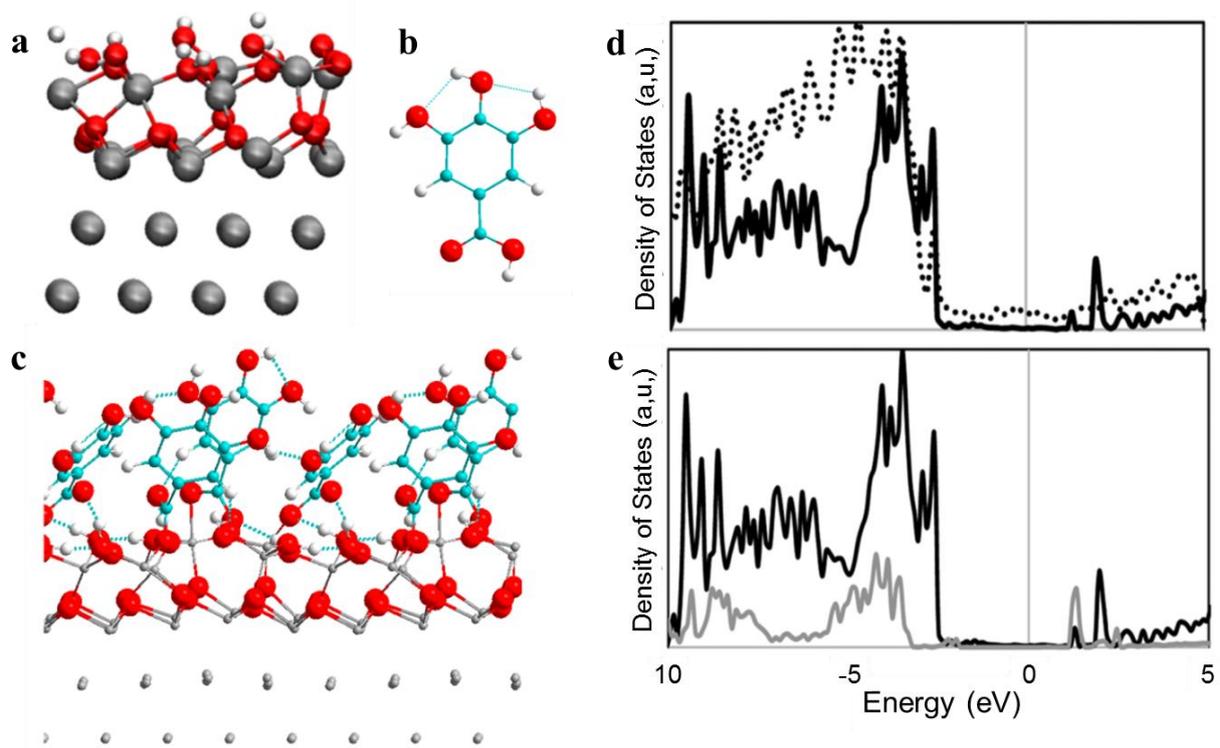